\documentclass[aps,prb,floatfix,twocolumn,showpacs]{revtex4}
\usepackage{graphicx}
\usepackage{dcolumn}
\usepackage{bm}
\newcommand{\figwidth}{3.5 in}
\begin{document}
\title{\bf String excitations of a hole
in a quantum antiferromagnet and photoelectron spectroscopy}
\author{Efstratios Manousakis}
\affiliation{Department of Physics and MARTECH, Florida State University, 
Tallahassee, FL 32306-4350, USA and\\
Department of  Physics, University of Athens, Panepistimiopolis, Zografos,
157 84 Athens, Greece.}  
\date{\today}
\begin{abstract}
The aim of the present paper is twofold. 
The first goal is to show that high resolution 
angle-resolved photoelectron spectra from cuprates
indicate the presence of string-like excitations of the quasihole excitation
in a quantum antiferromagnet.
In order to compare with the experimental intensity plots,
we calculate the spectral function of the $t-J$ and 
the $t-t^{\prime}-t^{\prime\prime}-J$ models within the self-consistent Born
approximation for widely accepted values of the parameters of these 
models. The main features of the high resolution photoelectron spectra
are in general agreement with the results based on the above models
and can be understood by considering not only the lowest
energy quasiparticle peak but, also, the higher energy string-like excitations
of the hole. These features are: (a) the energy-momentum  dispersion of 
the string-like excitations,  (b) the momentum dependence of the
spectral weight of the quasiparticle peak, and
(c) the way in which the spectral weight of the lowest energy
quasiparticle peak  is transfered to higher
energy string excitations, including the fact that it
vanishes near the $\Gamma$ point. 
The second goal of the present paper is to make the case that a 
proper analysis of both the numerical results obtained from these models and
of the experimental results, suggests a 
theoretical picture for the internal structure of the hole-quasiparticle and 
a {\it string-exchange  pairing mechanism} due to strong antiferromagnetic 
correlations among the background spins. We find that, 
using a simple model in which the Hilbert space is restricted to 
states of only unbroken strings attached to the holes and in which 
the holes are connected with unbroken strings, we can provide
a good quantitative description of the most accurate numerical
results. We find that the holes experience an effective interaction 
due to a string-exchange mechanism which is characterized by
 a rather large string tension and this can provide pairing energy scales 
much larger that those suggested by
spin-fluctuation mediated pairing models. In addition, it is argued that such 
string-exchange interaction
tends to bind holes in a bound state with the $d_{x^2-y^2}$ symmetry. 

\end{abstract}
\pacs{71.10.-w,71.10.Fd,71.27.+a,74.72.-h,79.60.-i}

\maketitle

\section{Introduction}
Angle-resolved photoelectron spectroscopy (ARPES) studies of the 
cuprates and related materials have been very useful in 
revealing the structure of the low energy
quasiparticle excitations\cite{wells,damascelli,ronning,graf}.
In this paper, we will focus on the most recent high resolution
ARPES studies\cite{ronning,graf} where an ``anomalous'' dispersion 
was identified
at relatively higher energy than the well-known low energy
quasiparticle band in the under-doped regime.
We would like to discuss the case of very light doping where
there are concrete predictions for the hole spectral function
in a quantum antiferromagnet\cite{liu,dagottor}. In this limit
the problem has been extensively studied and from that
ground and the above mentioned experimental studies a novel
physical picture emerges which consists of some already known features and 
some new ideas to be discussed here.

The motion of a single hole inside a quantum antiferromagnet
has been extensively studied using a number of analytical,
semi-analytical and numerical techniques. One of the
main and well-known conclusions of these studies
is that the hole becomes a rather well-defined low-energy quasiparticle
with the minimum of the quasiparticle
band at $(\pm\pi/2,\pm\pi/2)$, which is an unusual position in the Brillouin 
zone for a minimum to occur, and, thus, at sufficiently light
doping, a hole pocket develops  near $(\pm\pi/2,\pm\pi/2)$. 
In addition, quantum spin-fluctuations allow the
hole to move coherently as a polaron because of the fact that
they can erase the damage created by the hole and, vice versa, the
hole through its motion which displaces reversed spins can eliminate
some spin-deviations created in advance by quantum fluctuations; 
thus, the hole motion can lower the total energy because it repairs some 
damage by ``choosing'' to move through paths of already reversed spins.
Angle-resolved photoelectron studies of the undoped insulator
have demonstrated\cite{wells} that the minimum of the hole band is at
$(\pm\pi/2,\pm\pi/2)$ and the bandwidth is in very good agreement
to that predicted from the $t-J$ model\cite{liu} when the experimentally
determined value\cite{RMP} of $J\simeq 0.13 eV$ and the empirically
known value of $t \simeq 0.4 eV$ are used.  The dispersion
along the direction $(0,0)$ to $(\pi,\pi)$ as calculated
using the $t-J$ model\cite{liu} does not agree with that
measured by ARPES\cite{wells} and it requires an
extension of the $t-J$  model to add nearest neighbor ($t^{\prime}$)
 and next nearest neighbor ($t^{\prime\prime}$) hopping to
reproduce the experimental findings along this 
direction\cite{kim,theory,nazarenco,ferrell,lee,wheatley,belincher,eder}.

What is less well-known about the hole motion in a  
quantum antiferromagnet is what we call the string excitations
which are internal excitations of the hole polaron. 
These excitations arise because the hole along its path creates 
a string of spins which are reversed relative to the local 
antiferromagnetically correlated background. These excitations
have been discussed by Shraiman and Siggia\cite{shraiman0,shraiman1,
shraiman2} and investigated further by several 
authors\cite{barnes,dagotto,elser,liu}. 
In particular in Ref.~\onlinecite{liu}, where a detailed calculation
of the hole spectral function was reported, it was found that
these excitations appear as rather sharp peaks in the hole
spectral function and they compete with the low energy 
well-defined quasiparticle peak to gain spectral weight 
as a function of momentum.

The aim of the present paper is twofold.
One of the goals of the present paper is to discuss that the
recently reported high resolution angle resolved photoelectron 
studies from cuprates\cite{ronning,graf} strongly indicate the existence of 
such string excitations in these materials 
(See also, Ref.~\onlinecite{manousakis}).
A possible connection between such string 
excitations and certain features of the spectral function was 
made in past experimental studies\cite{ronning,kim}. In the present paper, 
a detailed analysis of their possible role in the spectral function
is provided along with a study of  
the consequences of their possible presence in doped quantum antiferromagnets. 
We provide a rather transparent physical
picture, through which we can understand several
features of the ARPES results in terms of the same unified framework.
The analysis provided in the present paper, by recalculating 
the hole spectral function using the $t-J$ model\cite{manousakis}
and  the $t-t^{\prime}-t^{\prime\prime}-J$ model
in order to properly compare the results with the experimental data, makes
a strong case for the existence of such string excitations
in the real materials.

The second important goal of the present paper is to present
a new framework
to understand the possible role of local strong antiferromagnetic
correlations in the doped cuprate materials in the pairing
mechanism. First, we introduce a model based on a reduced Hilbert
space which contains only unbroken strings of overturned
spins attached to the holes. It is demonstrated that this
reduced Hilbert space accurately captures the features
of the quasiparticle dispersion and the higher energy string
excitations. In addition, it is illustrated that using 
a subspace of states in which a pair of holes is connected with 
unbroken strings leads to a {\it string-mediated interaction} which
might give significantly stronger pairing mechanism than models
of pairing due to exchange of spin-fluctuations. 

The paper is structured as follows: In Sec.~\ref{calculation}
we present the results of the hole spectral function as calculated
using the $t-J$ and the $t-t^{\prime}-t^{\prime\prime}-J$ models
paying particular attention to the contribution and role of 
string excitations. In Sec.~\ref{comparison}, the analysis
of the recent ARPES experimental results in order to reveal
the contribution of the string excitations in the real materials
is made. Sec.~\ref{theory} presents the consequences of our
finding to our theoretical picture for the hole-quasiparticle
in doped cuprates. In addition, a simple model is presented
based on states of unbroken strings connected to the holes
which gives a very good account of the numerical results of the
$t-J$ model.  In addition, a pairing mechanism through string
exchange interaction is presented in the same section.

\section{\label{calculation} Calculation of the spectral function}

In this section of the paper we will use the same method used 
in Ref.~\onlinecite{liu} to calculate the spectral function
of a hole in a quantum antiferromagnet using the $t-J$ 
and $t-t^{\prime}-t^{\prime\prime}-J$ models with the aim to
make a direct comparison with the recent ARPES results.

In order to start the discussion, let us consider a 
quantum antiferromagnet doped with holes as described
by the well-known $t-J$ model
\begin{eqnarray} 
{\cal H} & = & -t \sum_{<ij>,\sigma} (c^{\dagger}_{i\sigma} 
c_{j\sigma} + c^{\dagger}_{i\sigma} c_{j\sigma})
+ J_z \sum_{<ij>} s^z_is^z_j \nonumber \\
& + & {{J_{\perp}}\over 2} \sum_{<ij>} 
(s^+_i s^-_j + s^-_i s^+_j)
\end{eqnarray}

which is restricted to operate in a single occupied subspace.
Here $c^{\dagger}_{i \sigma}$ creates a spin $\sigma$ electron 
at lattice site $i$. The operators $s^{x,y,z}_i$ are the
three components of a spin-1/2 operator and $s^{\pm}_i=s^x_i\pm i s^{y}_i$.
The first term takes into account the hole-hopping and the second
and third are the direct and exchange terms in the Heisenberg
antiferromagnet. Just the first two terms form a Hamiltonian
known as the  $t-J_z$ model. The $t-J$ model corresponds
to the spacial case when $J_z=J_{\perp}$.
 The $t-t^{\prime}-t^{\prime\prime}-J$ 
model is obtained from the above $t-J$ model by adding a
hopping term $t^{\prime}$  which allows the hole to move along
the diagonal of the square lattice and a $t^{\prime\prime}$ term
which moves the hole by two sites along the $ \hat x$  direction
(and $-\hat x$ direction) or by two sites along the $\hat y$ direction
(and $-\hat y$ direction).
The $t-t^{\prime}-t^{\prime\prime}-J$ model has been used in 
Refs.~\onlinecite{kim,theory,nazarenco,
ferrell,lee,wheatley,belincher,eder,avinash} to rather successfully
fit the photoemission data of Wells {\it et al.}\cite{wells}

In Ref.~\onlinecite{liu} following Ref.~\onlinecite{KLR}
an effective Hamiltonian was used where the Heisenberg spin interaction part
of the $t-J$ model was treated in spin-wave approximation and
the hole-hopping term was linearized in spin-deviation operators.
Within the non-crossing approximation, the Dyson's equation for the 
single-hole Green's function was solved self-consistently.
When the calculation was carried out on small-size lattices
a very good agreement was found with the exact diagonalization
results for the spectral function. However, the advantage of the
method is that it can be carried out on sufficiently
large size lattice where finite-size effects are entirely 
absent\cite{liu,marsiglio,martinez}.
Hence, we will use this method to calculate the hole spectral
function for the $t-J$  and $t-t^{\prime}-t^{\prime\prime}-J$ models 
with the aim to make a direct comparison with the ARPES results.

\begin{figure}[ht] 
\vskip 0.2 in
\begin{center}
  \includegraphics[width=3.0 in]{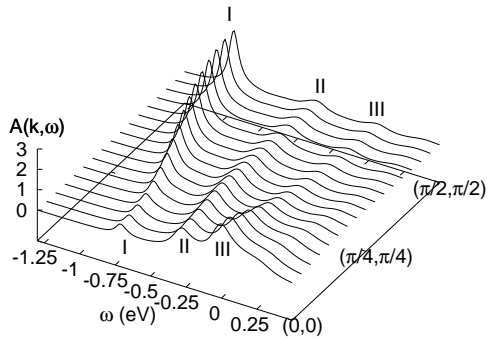}
\vskip 0.2 in
 \caption{\label{fig1} The spectral function of a single hole
in a quantum antiferromagnet for $J/t=0.3$ along the $(0,0)$ to 
$(\pi/2,\pi/2)$ direction as a function of energy $\omega$. }
\end{center}
\end{figure}

In Fig.~\ref{fig1} the hole spectral function is shown for the
$t-J$ model along the $(0,0)$ to $(\pi/2,\pi/2)$ direction
using the broadly accepted parameter values of 
$t=0.4 eV$ and $J/t=0.3$.
The Dyson's equation was solved as in Ref.~\onlinecite{liu}
using Lorentzian broadening with a width $\eta=0.1 t$. 
We note that while the width of the
first peak depends strongly on the value of $\eta$, as it is a
$\delta$ function peak, the other two (labeled II and III) and the 
higher peaks (not clearly seen in the graph due to the choice of the 
scale) remain unchanged
when we decrease the value of $\eta$. This point has been extensively
studied in  Ref.~\onlinecite{liu} and we have been able to reproduce it 
here. The first important point to be discussed in this paper
again is the higher energy peaks labeled II and III in the spectral
function. These peaks were studied in Ref.~\onlinecite{liu}
and it was concluded that they correspond to the string-like
excitations which can be understood in the simple $t-J_z$ model.
Before we go on in this paper we wish to briefly discuss the origin of these
peaks.

Let us first
imagine a  hole subject to quantum mechanics moving in a classical
antiferromagnetic template such as the one conceptualized within 
the $t-J_z$ model. As the hole tries to move, it creates
a string of spin deviations over the antiferromagnetic background
and, thus, it experiences a linearly rising 
potential\cite{shraiman0,shraiman1}.
As a result, if we ignore the small probability amplitude for
the hole to retrace its path\cite{trugman}, the hole is bound to 
its ``birth'' site. In addition, there are 
excited states\cite{barnes,dagotto,liu} of the
hole inside the almost linearly rising potential which in 
the limit $t/J>>1$ can be approximated by the Airy functions
and their corresponding eigenenergies are given by
\begin{eqnarray}
E_n/t=\epsilon_n + a_n (J_z/t)^{2/3}.
\label{peaks}
\end{eqnarray}

\begin{figure}[ht] 
\vskip 0.2 in
\begin{center}
  \includegraphics[width=3.0 in]{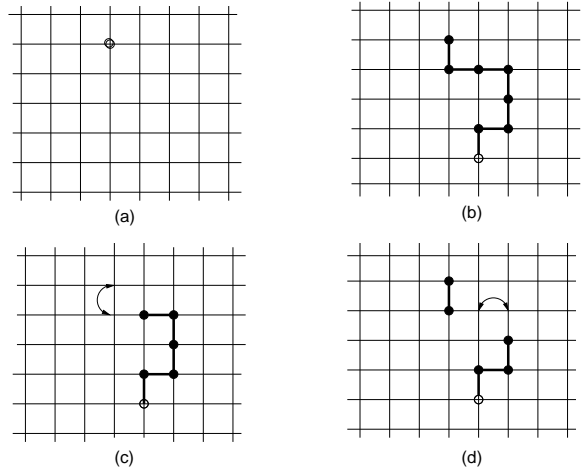}
\vskip 0.2 in
 \caption{\label{fig2} 
(a) A hole, shown by an open circle, in an antiferromagnetically aligned 
background (the spins of the N\'eel background are not shown in this
figure for clarity; only the spin-deviations are shown by a solid
circle) with
no spin-deviations. (b) The hole has moved along the shown path
by displacing the spin and creating the spin deviations shown with
the black dots and thus creating the string shown. (c) The 
pair-spin-flip term of the Heisenberg antiferromagnetic exchange
can eliminate a pair of nearest-neighbor spin deviation
at the beginning of the string and, thus, shortening the string
or (d) in the middle of the string thereby cutting the string
into a string with the hole at one end and just a string of spin-deviations.}
\end{center}
\end{figure}

Now, let us turn on the Heisenberg antiferromagnetic
exchange coupling $J_{\perp}$. This term creates or eliminates nearest-neighbor
pairs of spin-deviations and, thus, it makes it possible for
the hole to become free of its original ``birth'' site and find
its ``umbilical-cord'' attached to a new ``birth'' origin.
As shown in Fig.~\ref{fig2} this can either happen by eliminating
the pair of spin-deviation at the beginning of the string
involving the ``birth'' location of the
hole (Fig.~\ref{fig2}(c)  or anywhere along the string (Fig.~\ref{fig2}(d), 
thus, producing two strings,
one attached to the hole and the other becomes part of the 
background spin-deviations, i.e., part of the fluid of the
spin-deviations which always exist in a quantum antiferromagnet
due to zero temperature quantum fluctuations. In addition, to these
processes there can be processes where the creation of pairs of
spin-deviations adjacent to the hole position precedes the hole motion, 
in which case the hole experiences the string as a line of downhill
potential. In the limit of $t/J_{\perp}>>1$, however, the
spin-relaxation time is much longer that the characteristic time
for hole-hopping. Thinking in the same spirit as the
Born-Oppenheimer approximation for the electronic motion
relative to the slowly moving ions, one can convince himself that
the approximate notion of a linearly rising potential exists.
The important difference from the simpler $t-J_z$ model
is that these spin fluctuations, no matter how slow they maybe, play the
leading role in delocalizing the ``birth'' site from where
the string is attached.
This process is the process which gives rise to the hole band
with a minimum at $(\pi/2,\pi/2)$. We would like
to stress, however, that a more correct qualitative picture is
one where the hole is held by a string from a mobile birth-site.    
We will come back to this discussion in Sec.~\ref{theory} were
we will examine further the role of these processes.

 In Ref.~\onlinecite{liu}, the peaks II and III in Fig.~\ref{fig1}, 
have been attributed to the internal excitations of the string
analogous to the excitations of a particle in a linear potential.
The energy where the peaks are located in the spectral function
have been studied as a function of $J/t$ (in the $t-J$ model
where $J_{\perp}=J_z=J$) and it was found that they fit
to the form given by Eq.~\ref{peaks} with $J_z$ replaced by
$J$ in a wide range of $J/t$. In addition, the width of these
peaks was found to scale with $J/t$.
In Sec.~\ref{theory}, the emergence of these peaks as higher energy
excitations of the hole restricted in a Hilbert space of states 
made of only unbroken strings is discussed.

In the present paper, one of the important aspects of Fig.~\ref{fig1} 
that we notice is the
vanishing of the spectral weight of the low energy quasiparticle
peak as we move away from the minimum of the band at $(\pi/2,\pi/2)$
toward the $\Gamma$ (momentum $(0,0)$) point. In addition, notice
that the spectral weight is transfered to the higher energy
peaks which correspond to the string resonance excitations.

\begin{figure}[ht] 
\vskip 0.2 in
\begin{center}
  \includegraphics[width=3.0 in]{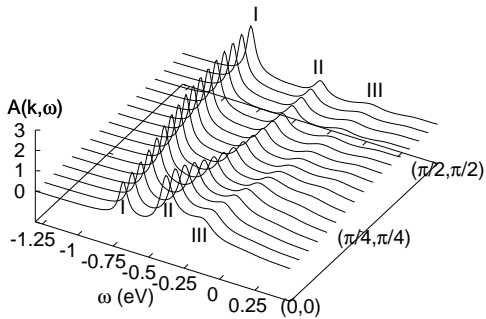}
\vskip 0.2 in
 \caption{\label{fig3} The spectral function of a single hole
in a quantum antiferromagnet for the $t-t^{\prime}-t^{\prime\prime}-J$
model with the same parameter values used in Ref.~\onlinecite{kim}
 along the $(0,0)$ to 
$(\pi/2,\pi/2)$ direction as a function of energy $\omega$.  }
\end{center}
\end{figure}

In Fig.~\ref{fig3} the calculated hole spectral function 
for the $t-t^{\prime}-t^{\prime\prime}-J$
model with the same parameter values used in Ref.~\onlinecite{kim} 
($t=0.35 eV$, $t^{\prime}=-0.12 eV$, $t^{\prime\prime}=0.08 eV$ and
$J=0.14 eV$) is shown along the $(0,0)$ to 
$(\pi/2,\pi/2)$ cut. Again, in this case also, the
Dyson's equation has been solved as in Ref.~\onlinecite{liu}
using Lorentzian broadening with a width $\eta=0.1 t$.
This figure should be compared with Fig.~\ref{fig1} obtained for the
pure $t-J$ model. All important aspects of the spectral function
obtained with the simpler $t-J$ model remain the same.
The only difference is that as we approach
the $\Gamma$ point more spectral weight  is transfered to the
peak II instead of III.
\begin{figure}[ht] 
\vskip 0.2 in
\begin{center}
  \includegraphics[width=3.0 in]{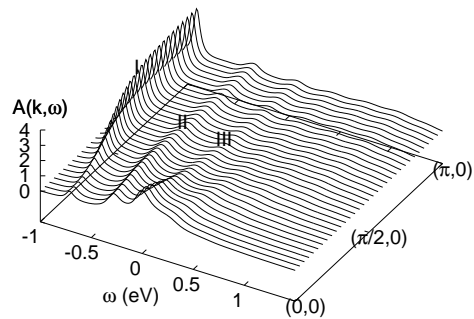}
\vskip 0.2 in
 \caption{\label{fig4} The spectral function of a single hole
in a quantum antiferromagnet for the $t-J$
model along the $(0,0)$ to 
$(\pi,0)$ direction as a function of energy $\omega$. }
\end{center}
\end{figure}

\begin{figure}[ht] 
\vskip 0.2 in
\begin{center}
  \includegraphics[width=3.0 in]{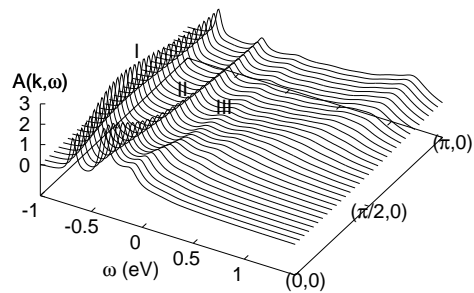}
\vskip 0.2 in
 \caption{\label{fig5} The spectral function of a single hole
in a quantum antiferromagnet for the $t-t^{\prime}-t^{\prime\prime}-J$
model with the parameters used in Ref.~\onlinecite{kim}
($t=0.35 eV$, $t^{\prime}=-0.12 eV$, $t^{\prime\prime}=0.08 eV$ and
$J=0.14 eV$)  along the $(0,0)$ to 
$(\pi,0)$ direction as a function of energy $\omega$. }
\end{center}
\end{figure}

In Fig.~\ref{fig4} and Fig.~\ref{fig5}, the calculated hole spectral 
function along the $(0,0)$ to $(\pi,0)$ cut is shown for the
$t-J$ model and the $t-t^{\prime}-t^{\prime\prime}-J$
model respectively. The low energy quasiparticle excitation
along this direction attains it minimum at $(\pi,0)$ in the 
$t-J$ model (Fig.~\ref{fig4}) where the peak acquires most of the spectral
weight. This is not the case in the experiment. As a result
the $t-J$ model was extended to the $t-t^{\prime}-t^{\prime\prime}-J$
model  precisely  for this reason. Notice in Fig.~\ref{fig5}
that the band minimum along this direction is no longer at
$(\pi,0)$ but rather at $(\pi,/2,0)$ where the spectral weight in
maximum. Again, notice that the spectral weight as we approach the
$\Gamma$ point is transfered to the second peak due to the internal
excitation of the string.

\section{\label{comparison}
Comparison with experiment}

\label{experiment}
\begin{figure}[ht] 
\begin{center}
\includegraphics[width=\figwidth]{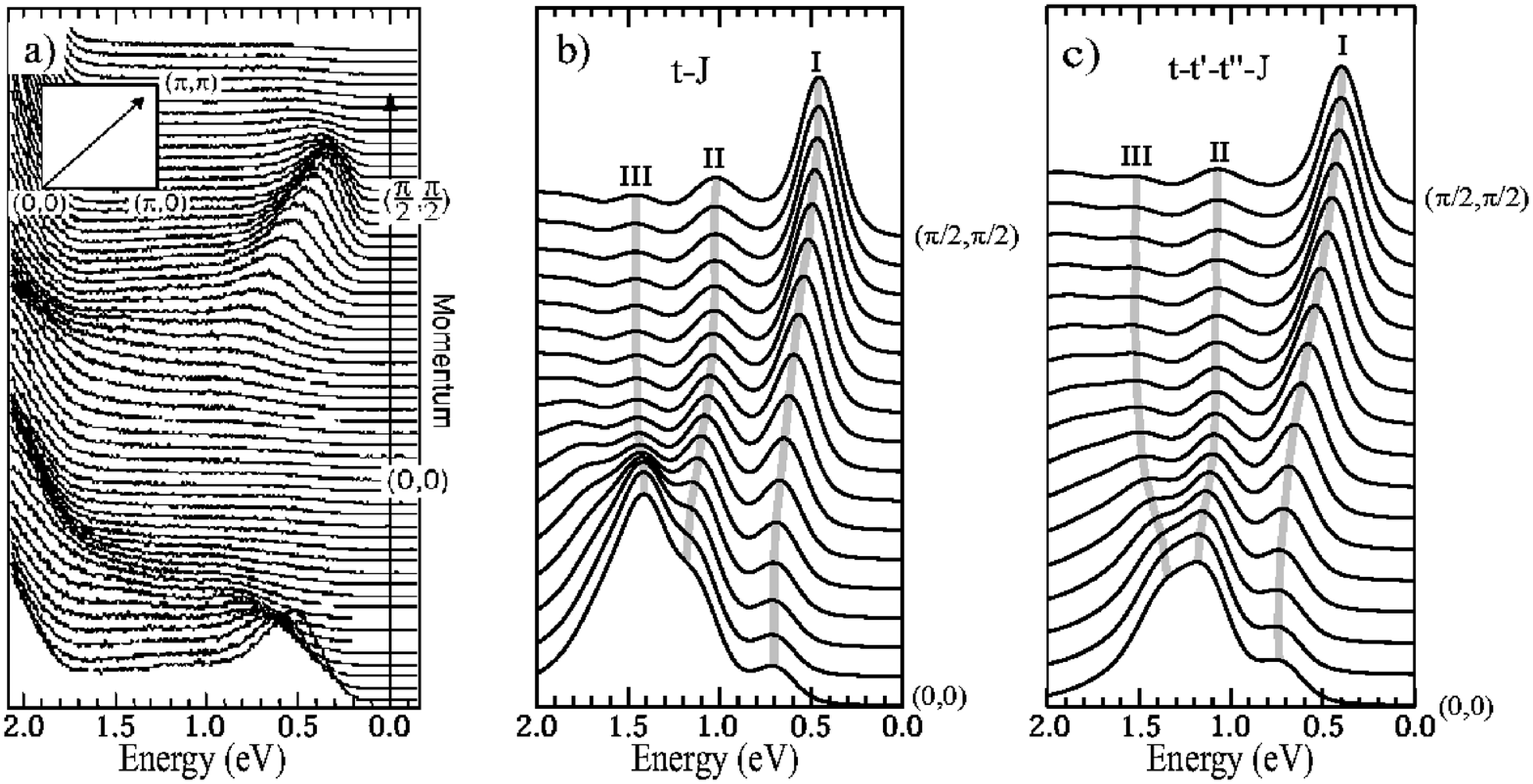}
 \caption{\label{fig6} 
The experimentally determined spectral function
with high resolution ARPES\cite{ronning} shown on the left (part a) 
is compared with  the theoretical spectral functions obtained
from the $t-J$ model\cite{liu} for $J/t=0.3$ and $t=0.4 eV$ along the path from
$(0,0) \to (\pi/2,\pi/2)$ shown in the middle (part b), and 
 that obtained from the $t-t^{\prime}-t^{\prime\prime}-J$ model (part c) 
along the  previously mentioned Brillouin Zone path.
The solf-gray lines labeled I,II and III are guides to the eye. 
The reference energy has been shifted by a 
constant so that the energy of the quasiparticle peak at $(\pi/2,\pi/2)$ to
be the same as the experiment.}
\end{center}
\end{figure}

The lowest energy peak at $(\pi/2,\pi/2)$ in the theoretical calculation 
is very sharp while in the experimental data it carries a width of the 
order of a fraction of an $eV$. It is not clear what is the exact mechanism
which causes the damping of the quasiparticle and there are reasons
to believe that it may be due to the coupling to phonon 
excitations\cite{nagaosa}. Therefore, in order to compare with the
experimental results we need to broaden the results of our
calculation. We use a Gaussian broadening function, namely 
$A_b({\bf k},\omega)= \int A({\bf k},\omega') 
G(\omega-\omega') d\omega'$, where
$G(\omega-\omega') = exp(-(\omega-\omega')^2/\epsilon^2)$, using 
$\epsilon = \sqrt{2} \sigma = 0.125 eV$. 
We have obtained very similar results when
we use a value of $\eta=0.125 eV$ in the initial propagator 
namely in the iterative procedure to solve the Dyson's equation.
 The same amount of broadening has been used in Ref.~\onlinecite{kim} 
to compare the ARPES peak to the quasiparticle peak given by 
the  $t-t^{\prime}-t^{\prime\prime}-J$ model.

\begin{figure}[ht] 
\begin{center}
\includegraphics[width=\figwidth]{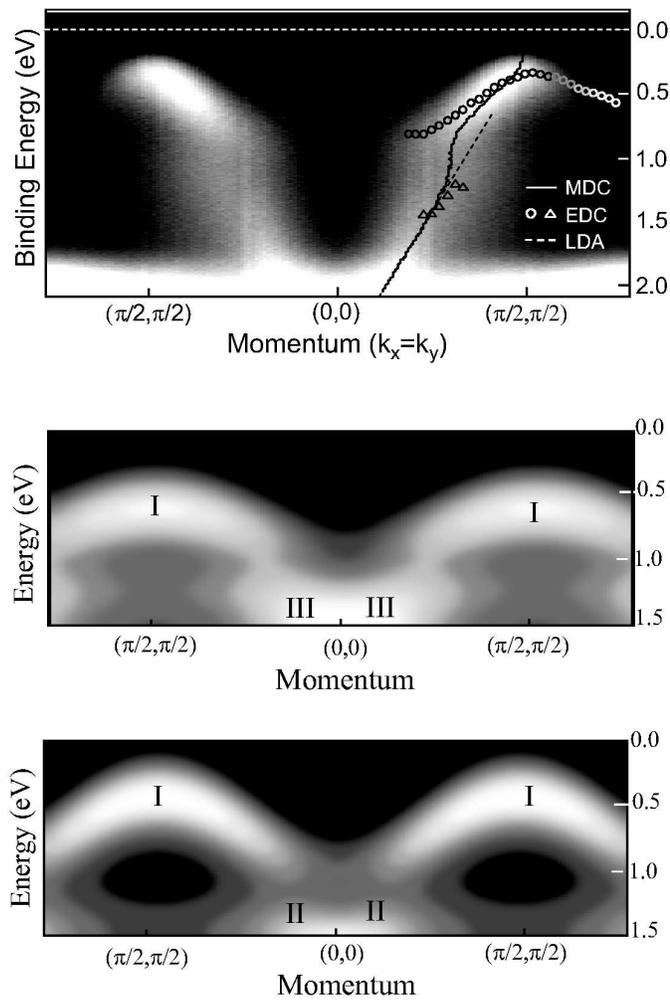}
 \caption{\label{fig7} 
The experimentally observed intensity 
plot reported in Fig.~2 of Ref.~\onlinecite{ronning} (top) is compared with
that obtained from the $t-J$ model (middle) and from the 
$t-t^{\prime}-t^{\prime\prime}-J$ model (bottom). 
The reference energy has been shifted by a 
constant so that the energy of the quasiparticle peak at $(\pi/2,\pi/2)$ to
be the same as the experiment. }
\end{center}
\end{figure}

\begin{figure}[ht] 
\begin{center}
\includegraphics[width=\figwidth]{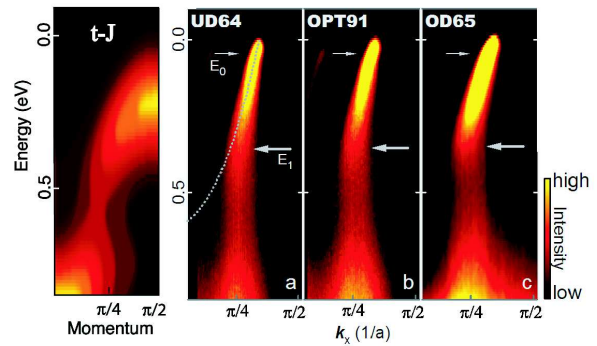}
 \caption{\label{fig8} 
Left: The theoretical color coded intensity plot along 
the cut indicated in the top-right inset, as obtained from the $t-J$ model
using the widely acceptable values for the parameters $J/t=0.3$ and $t=0.4 eV$.
Right: Three ARPES intensity plots for underdoped, optimally doped
and overdoped $Bi2212$ taken from Fig.~1 
of Ref.~\onlinecite{graf} for comparison.}
\end{center}
\end{figure}

In Fig.~\ref{fig6} a comparison is presented between the experimental
spectral function along the $(0,0)$ to $(\pi,\pi)$ cut, 
as reported in Ref.~\onlinecite{ronning} by means
of high resolution ARPES (left or part (a)), and those obtained 
from the $t-J$ model (part (b)) and the
 $t-t^{\prime}-t^{\prime\prime}-J$ model (part (c)). We would like
to notice that both in the theoretical and the experimental results,
 as we move away from $(\pi/2,\pi/2)$ toward
$(0,0)$, the spectral weight of the low energy well-defined quasiparticle peak
vanishes and it is gradually transfered to higher energy excitations. 
Notice that in the comparison of Fig.~\ref{fig6} and in the comparison
provided in Fig.~\ref{fig7} and Fig.~\ref{fig8}, if the energy scale
of the theoretical results (i.e., the value of $t$) is increased 
by about $25\%$ (such that
the theoretical peak at the $\Gamma$ point at  $\sim 1.5 eV$ is 
pushed to  $\sim 2 eV$) the agreement
between theory and experiment becomes better, especially
at higher energies near $2 eV$. However, at such high
energy scales we expect other effects to become important
and the description, in terms of a Zhang-Rice\cite{ZR} singlet within
the simple $t-J$ model, is also expected to fail.  

In Fig.~\ref{fig7}, we have produced an intensity plot
along the $(0,0)$ to $(\pi,\pi)$ cut, in order to compare directly
our results with those of Ref.~\onlinecite{ronning}.
For the purpose of comparison,
 we have added the experimental results
of Ref.~\onlinecite{ronning} as reported in their Fig.~2
(top of Fig.~\ref{fig7}).
In the second intensity plot from the top, we present that
obtained from the $t-J$ model with the generally accepted 
parameter values mentioned earlier. The bottom intensity
plot is that obtained from $t-t^{\prime}-t^{\prime\prime}-J$ model
using the parameter values used in Ref.~\onlinecite{kim}.
The intense peaks at $(\pi/2,\pi/2)$ and 
near $(0,0)$ are present in both theoretical 
and experimental intensity plots.
As we approach  momentum $(0,0)$ the spectral weight is
transfered to higher energy ``string'' states (II and III) as also
was discussed in Sec.~\ref{calculation}. This
gradual transfer manifests itself as a more luminous path
connecting the bright peaks at $(\pi/2,\pi/2)$ and $(0,0)$.

In a very recent  ARPES study\cite{graf} an anomalous
 dispersion was identified and a second energy  scale at around $0.8 eV$ 
was reported.  
In our calculation this energy
scale is the average energy of the string excitation 
peaks II and  III (measured from the lowest energy state at 
$(\pi/2,\pi/2)$). This becomes evident by comparing 
our Fig.~\ref{fig7} with Fig.~2 of Ref.~\onlinecite{graf}.
In Fig.~\ref{fig8} we present
a theoretical color-coded intensity plot (left) along 
the $(\pi/2,\pi/2)$ to $(0,0)$ cut, as obtained from the $t-J$ model
to be compared with the three ARPES intensity plots (right part of
Fig.~\ref{fig8}) for underdoped,  optimally doped
and overdoped $Bi2212$ taken from in Fig.~1 
of Ref.~\onlinecite{graf}. A very similar intensity plot
was obtained for the case of the $t-t^{\prime}-t^{\prime\prime}-J$ model. 
First of all in making a comparison we must realize that
the lowest energy quasiparticle states,  which correspond
to the lowest energy string states in the neighborhood 
of $(\pi/2,\pi/2)$, are occupied by holes which form a
Fermi sea. Namely, these occupied states will be entirely dark in
the intensity plot and the  states with highest intensity would be right
on the intersection of the Fermi surface with the most luminous
states of the theoretical intensity plot.  
Therefore, by assuming these adjustments
the color-coded intensity plot of Fig.~\ref{fig8}, excluding the bright
spots around $(\pi/2,\pi/2)$ (because at sufficient amount of doping the
states around $(\pi/2,\pi/2)$  should be inside the Fermi sea and the 
bright spots  should move just outside the Fermi surface), qualitatively
agrees with the experimental intensity plots reported in 
Fig.~1 of Ref.~\onlinecite{graf}.
These figures and the previous discussion clearly indicate 
that the process of  spectral weight  transfer to the higher
string states is masked by low intensity and/or broadening and
manifests itself in the intensity plot as an ``anomalous'' dispersion with
the center of the anomaly close to $(\pi/4,\pi/4)$ as in Fig.~1 of
Ref.~\onlinecite{graf}. We find it very interesting
that these features persist all the way up to the overdoped regime.
Our conclusion, therefore, is  that the spin correlations
should be strongly antiferromagnetic even in the overdoped
cuprates and this point is discussed in the next section. 

\section{\label{theory}
The string mediated interaction}

\subsection{Role of string states in the single-hole dispersion}

As we discussed the origin of the spectral function peaks
is due to the string excitations and they provide
a qualitative explanation of the
vanishing of the spectral weight from the low energy quasiparticle
peak by means of spectral weight shift to higher energy
string excitations. In this part of the paper, motivated by these 
finding we will consider 
a new theoretical approach to understand doped quantum antiferromagnets,
one in which such string excitations play a fundamental role.

First we notice that as the hole moves it does not simply 
create spin deviations, it creates (or destroys) a spatially connected 
string of spin-deviations. Therefore, as a basis to span the Hilbert 
space instead of disconnected spin-deviations we will consider an 
over-complete basis which is spanned by all possible strings which
define connected paths of spin-deviations as shown in Fig.~\ref{fig2}.
The spin fluctuation part
of our Hamiltonian conserves spin and creates or annihilates pairs
of spin deviations which can be regarded as strings of two sites. 
The only case in which a single spin-deviation
is needed is when we have a state where a spin-deviation is next to 
the site where the hole is located. 
This state can be regarded as a hole attached
to a string of length unity.

The fundamental reason for the choice of this basis is that
in the limit where $t>>J$, the quantum antiferromagnet with
just a small hole concentration should look like a fluid of strings
of overturn spins created by the motion of the holes with 
a lot of nearest-neighbor pairs of spin deviations (due to the
spin conserving antiferromagnetic exchange) which can also be
considered as strings of size two. In this regime, the 
average string size  in the ground state should be
larger than two. In addition, this basis describes better the short
wavelength physics and it might provide a better description
in the doped regime when the long-range spin correlations
are destroyed, but the relatively short-range antiferromagnetic
correlations are still present.

\begin{figure}[ht] 
\vskip 0.2 in
\begin{center}
  \includegraphics[width=3.0 in]{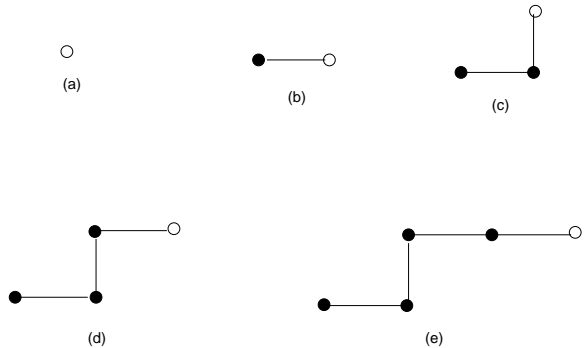}
\vskip 0.2 in
 \caption{\label{fig9} 
Connected single-hole string states up to length 4. The hole
is represented by the open circle and the filled circle 
stands for a spin deviation.
(a) The hole in the N\'eel background. (b) There are 4 states
of string length unity. (c) There are 12 states of string length  2.
(d) 36 states having string length 3 exist. (e) There are 
108 states such the ones shown in Fig.~(e) above with strings of
length 4.}
\end{center}
\end{figure}
\begin{figure}[ht] 
\vskip 0.2 in
\begin{center}
  \includegraphics[width=3.0 in]{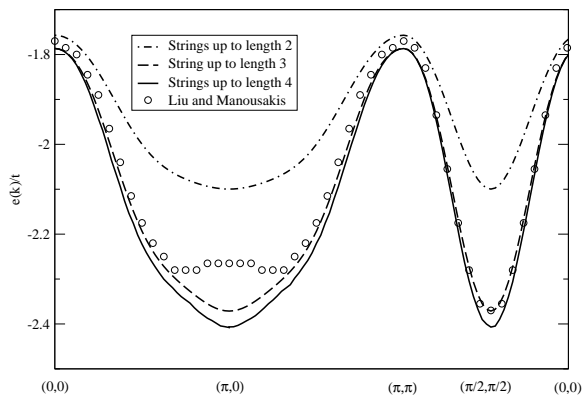}
\vskip 0.2 in
 \caption{\label{fig10} 
The hole dispersion as calculated with a Hilbert space made of only
strings. The dispersion is shown along the Brillouin zone path 
$(0,0)\to (\pi,0) \to (\pi,\pi) \to (0,0)$.
The three dispersion relations are calculated with 
string states of up to a limiting length size which is indicated
in the figure. The result is compared to those obtained from
the full calculation of the dispersion curve from the spectral
function discussed in Sec.~\ref{calculation}}
\end{center}
\end{figure}

To show the importance of string excitations, we diagonalized the
$t-J$ Hamiltonian in a restricted subspace of only states having a 
string connected
to the hole. These states are shown in Fig.~\ref{fig9}(a-e). They
can be characterized by the length of the string and we have successively
included all such states with strings up to length 4. In Fig.~\ref{fig10}
we present the calculated dispersion relations where the energy
is obtained by including strings states of (a) up to string length
2 (dashed-dot solid line) (b) up to string length 3 (dashed curve) and (c) 
up to length 4 (solid curve).  Notice that
the agreement with the results obtained with the method discussed
in Sec.~\ref{calculation} is remarkably close. The absolute 
energy scale at $k=0$ is accurately reproduced and, in addition, 
notice the close agreement of the dispersion near $(\pi/2,\pi/2)$. There is
some disagreement near $(\pi,0)$. However, this part of 
the dispersion is not described accurately by the full 
solution to the $t-J$ model either and we needed to introduce
the $t^{\prime}$ and $t^{\prime\prime}$ terms which 
give the leading correction to the dispersion near this point. 

The previous calculation suggests the following variational single-hole 
wavefunction, where the effect of background spin-fluctuations above
the N\'eel state is turned on:
\begin{eqnarray}
| \psi_{\bf k} \rangle  & = & 
  { 1 \over \sqrt{N}} \sum_{\bf R} e^{-i {\bf k} \cdot
{\bf R}} \hat T_{\bf R} \sum_{\sigma}c_{{\bf R}\sigma}| 0 \rangle,  \\
\label{singlehole}
\hat T_{\bf R} & = & 1 + \sum_{m=1}^n 
\sum^{\prime}_{\{\vec \delta_i\}} 
\Psi^m_{\bf R}(\{\vec \delta_i \}) 
\hat T^m_{\bf R}(\{\vec \delta_i\}),
\label{string}
\end{eqnarray}
where
\begin{eqnarray}
\hat T^m_{\bf R}(\{\vec \delta_i\}) & = & 
\hat P_{{\bf R}+\vec \delta_{m-1},\vec \delta_m} ...
\hat P_{{\bf R}+\vec \delta_1,\vec \delta_2} P_{{\bf R},\vec \delta_1},
\label{stringop}
\end{eqnarray}
where  
$\hat P_{{\bf R'},\delta} =  \sum_{\sigma} c^{\dagger}_{{\bf R'},\sigma}
c_{{\bf R'}+\vec \delta,\sigma}$ is the hole-hopping operator of
the $t-J$ Hamiltonian which creates the strings and
$\Psi^m_{\bf R}(\{ \vec \delta_i \})$ are variational parameters.
The operators $\hat T^m_{\bf R}(\{\vec \delta_i\})$ create strings
of overturned spins of length $m$ with the hole at ${\bf R}$
along the path ${\bf R} \to {\bf R} + \vec \delta_1 \to {\bf R} + \vec 
\delta_1+\vec \delta_2 \to ... \to {\bf R} + \vec 
\delta_1+\vec \delta_2 + ... + \vec \delta_m$.
The prime on the summation means that it is over paths such
as $\vec \delta_{i+1} \ne -\vec \delta_i$.
The summation over $m$ is over states of strings of length $m$ (up to $n$) 
connected with the hole. The following state can
be used for the state $| 0 \rangle$ in order to include the effect of
 spin-fluctuations
\begin{eqnarray}
| 0 \rangle & = & \sum_{\alpha} (-1)^{L(\alpha)} \exp \Bigl (-{1 \over 2} 
\sum_{i<j} u_{ij} S^{z}_i S^{z}_j \Bigr ) | \alpha \rangle,
\label{nohole}
\end{eqnarray}
where the sum is over all spin configurations and $L(\alpha)$ is the
so-called Marshall phase\cite{RMP}, namely $L(\alpha)$ is the number of 
up spins in one sublattice in the configuration $\alpha$.
When $u_{ij}=0$ the state $| 0\rangle$ is the N\'eel state
along the x-direction in the spin space.
 The variational wavefunction given by Eq.~\ref{singlehole} when $u_{ij}=0$
is identical to that obtained from the diagonalization discussed previously 
(with results given in  Fig.~\ref{fig10}).
Furthermore the function $u_{ij}$ is the spin-spin correlation
function and describes spin-fluctuations\cite{RMP} around the N\'eel ordered
state.
Variational calculations for the Heisenberg antiferromagnet
using the no-hole state given by Eq.~\ref{nohole} give 
very accurate results for
the ground state energy, the staggered magnetization and
the excitations above the ground state via sum rules\cite{RMP}.
In Ref.~\onlinecite{BM1} the above wavefunction with $n=2$ and 
$u_{ij}$ the same as the one which describes the no-hole
case accurately\cite{liu2} was used to carried a variational 
Monte Carlo calculation.  The results of this calculation 
for the single hole dispersion were accurate for $J/t>1$.
For $J/t\sim 0.3$, longer string states were required ($n>2$), which 
is a conclusion
very similar to the present findings presented in Fig.~\ref{fig10}.
These longer string excitations were allowed in the Green's function Monte
Carlo(GFMC) calculation of Ref.~\onlinecite{BM1b} which iteratively 
projects the correct single hole state starting from such
a wave function as an initial state. This calculation
converges to the correct energy with only a few GFMC iterations 
which indicates that
this wavefunction is a good starting point having relatively large
overlap with the exact wavefunction. 

\begin{figure}[ht] 
\vskip 0.5 in
\begin{center}
  \includegraphics[width=3.0 in]{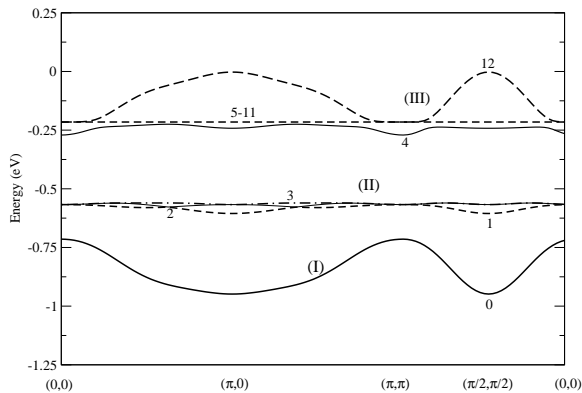}
 \caption{\label{fig11} 
The higher energy string excitations obtained by diagonalizing
the $t-J$ Hamiltonian in the subspace spanned by the 161 states
illustrated in Fig.~\ref{fig9} using $J/t=0.3$ and $t=0.4 eV$. 
The lowest energy state
is labeled 0 and it is the same as the dispersion presented in 
Fig.~\ref{fig10}. The lowest lying excited states are three nearly
degenerate excited states labeled 1,2,3 (which form the group labeled (II)). 
Following a gap there is a group of states labeled (III) which includes
the fourth state and the next six states, namely, 5-11 which are degenerate.
In addition, it includes the 12th excited state which follows a 
mirror-like dispersion to the ground state.  }
\end{center}
\end{figure}

In Fig.~\ref{fig11} we give the energy of the string excitations found 
by diagonalizing the $t-J$ Hamiltonian in the subspace of 
the string states shown in Fig.~\ref{fig9} along the same
Brillouin zone path, namely, from $(0,0)\to (\pi,0) \to (\pi,\pi) \to (0,0)$.
These states form bands and we can identify what we have labeled
as II and III in the previous discussion of Sec.~\ref{calculation}
and Sec.~\ref{comparison}. Namely, the three nearly degenerate excited
states labeled as 1,2 and 3 form the band II and the nine states 
labeled 4-12 form the group labeled III in our previous discussion of 
Sec.~\ref{calculation}.
Notice that their energy is the same to those of the peaks labeled
II and III in Fig.~\ref{fig1} and Fig.~\ref{fig4}. In addition, their
dispersion is very similar, namely, II is  almost dispersionless
and III along the $(0,0)$ to $(\pi/2,\pi/2)$ cut (Fig.~\ref{fig1})
moves closer to the peaks I and II near $(0,0)$
and further away from the peaks I and II near $(\pi/2,\pi/2)$.

Therefore, the conclusion of this study is that the short-range
string correlations are very important in determining the energetics
of the hole dynamics. One expects that long wavelength spin
fluctuations to be important in the ordered quantum antiferromagnet.
However, when the antiferromagnetic order
is destroyed as the doping level is increased, while the long wavelength 
spin-fluctuation
excitations are not well-defined, the string excitations still remain
and they can quantitatively account for the short wavelength
dynamics. 
With these findings in mind we proceed to show that these
string correlations give rise to a  pairing mechanism which is 
characterized by a relatively strong pairing interaction,
namely, significantly stronger that the effective interaction
due to exchange of antiferromagnetic spin-fluctuations.

\subsection{String-mediated pairing} 
In this paper, inspired by the findings of the photoemission measurements
and our suggestion that they provide evidence for the existence
of string excitations even at optimal doping, we suggest that
the pairing mechanism found numerically in the $t-J$ model\cite{BM,KM}
is mediated by strings. 

\begin{figure}[ht] 
\vskip 0.2 in
\begin{center}
  \includegraphics[width=3.0 in]{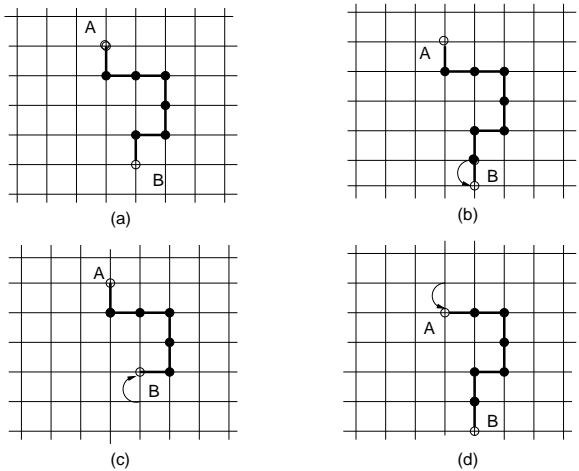}
\vskip 0.2 in
 \caption{\label{fig12} 
(a) A pair of holes (A) and (B) marked by o's are connected by a 
string of overturn spins. (b) For either hole to attempt to unbind itself
from the other leads to lengthening the string which is energetically costly.
(c) For either hole to attempt to hop closer to the other
hole it is energetically favorable.
(d) The string configuration of pair of holes (a) can be momentarily
lengthened by allowing hole B to hop as shown in (b) and finally
the length of the string is restored to original length size
by allowing hole A to hop along the string as in configuration (d). }
\end{center}
\end{figure}

The rather simple calculation of the previous
subsection suggests that states of unbroken strings 
are the main contributors to the single-hole dispersion.
In Fig.~\ref{fig12} a pair of holes A and B 
is shown connected by a string of overturned spins in an otherwise
idealized N\'eel ordered background with no spin deviation.
First, in order to facilitate the discussion let us consider
states with unbroken strings connecting the two holes. 
The two holes experience a potential linearly rising
as a function of the length of the string between them.
 Namely, when either hole is moved in such a way that the
distance between the holes is increased as shown in Fig.~\ref{fig12}(b), 
it stretches
the length of the string of overturned spins and the potential
energy increases. On the other hand, each hole feels a 
{\it down-hill potential}
toward each other because by moving along the path, defined by
their mutually connecting string, they repair the overturned
 spins as indicated in Fig.~\ref{fig12}(c).

In addition, it can be demonstrated using Fig.~\ref{fig12} (a,b,d)
that when there are such string
correlations between two holes, the entire string bipolaron can
move through the lattice with an effective mass which is lighter than
that of a single hole. In leading order the hole has to make two 
successive hops  followed by a pair spin-flip 
in order for the hole to move
through the lattice. This is of higher order than the process 
illustrated in Fig.~\ref{fig12}(a,b,d) where first the hole A hops
(transition from (a) to (b)) momentarily lengthening the string
and this is followed by a hop of the second hole B
(transition from (b) to (d)) which takes the down-hill path
formed by the string. Notice that the configuration (d) has the same
length of the string as the starting configuration A, but the hole 
string bipolaron has moved and, thus, gains {\it delocalization energy}.


A variational wave function to describe such a pair of holes 
of zero net momentum may be written as follows:
\begin{eqnarray}
|\psi_2 \rangle &=&
\sum_{{\bf r}_1,{\bf r}_2} e^{- i {\bf k} \cdot ({\bf r}_1-{\bf r}_2)}
g({\bf r}_2-{\bf r}_1) \nonumber \\
&\times &\Bigl [1 + \sum_{{\bf s}({\bf r}_1,{\bf r}_2)} 
\Phi(Y({\bf r}_1,{\bf r}_2)) \hat Y({\bf r}_1,{\bf r}_2)\Bigr ] 
\hat T_{{\bf r}_1} \hat T_{{\bf r}_2}
 | 0 \rangle,
\label{variational}
\end{eqnarray}
where $Y({\bf r}_1,{\bf r}_2)$ is a connected path connecting the sites
${\bf r}_1$ and ${\bf r}_2$ which are the location of the holes and 
$\Phi(Y({\bf r}_1,{\bf r}_2)))$  is a function of the path 
which defines the string
and it is expected to be a decreasing function of the length of the
path. The state $| 0 \rangle$ is the no-hole state given by Eq.~\ref{nohole}
and the operators $T_{{\bf r}_{1,2}}$ are given by Eq.~\ref{string}
and they create the strings attached to each hole.
The operator
$\hat Y({\bf r}_1,{\bf r}_2)$ is a product of hole-hop operators
$\hat P_{ij}$ (defined after Eq.~\ref{stringop}) along the path 
connecting the two holes defined by   $Y({\bf r}_1,{\bf r}_2)$. 
When this operator  acts on the N\'eel state creates states, such 
as the  state illustrated in Fig.~\ref{fig12}(a), where the holes
located at ${\bf r}_1$ and ${\bf r}_2$ are connected with a
string of overturned spins in an otherwise perfectly ordered
N\'eel spin lattice. 
The function $g({\bf r}_2-{\bf r}_1)$ gives the symmetry
of the two hole state and for $d_{x^2-y^2}$ symmetry it 
changes only its sign (keeping the same magnitude) for a $\pi/2$ rotation.  

A similar  variational state built upon such a basis of string states
which included the effect of background spin fluctuations
was studied in Ref.~\onlinecite{BM} and it was also used as
the initial state for the Green's function Monte Carlo
calculation. It was found that  two holes bind in a state with
the $d_{x^2-y^2}$  symmetry for $J/t$ greater than about $0.27$.

Let us restrict ourselves in the subspace where the two 
holes are connected by an unbroken string. 
This implies that the spin-fluctuation part of the
$t-J$ model gives zero contribution to matrix elements
in this subspace. In order to illustrate that, let us consider 
a two-hole state with an unbroken string mediated between the two holes
such as any state from Fig.~\ref{fig12}. 
When applying the pair spin-flip operator of the $t-J$ model
on such a state, it will always produce a state with a broken 
string and, thus, it takes
us outside the Hilbert space of states with all-connected strings.
As a consequence the result of the calculation of the binding energy $\Delta$ 
of two holes in this restricted subspace,
is very similar to that using the $t-J_z$ model with $J_z=J$.
Using a diagrammatic approach Chernyshev and 
Leung\cite{chernyshev} found that there is strong binding
in $p$ and $d$ wave symmetry between two holes in the $t-J_z$ model. 
In addition, calculation of the quantity $\Delta$ has been carried out 
for the $t-J_z$ model using exact 
diagonalization\cite{riera} on lattices with up to 50 sites and 
it was found that the two holes bind in a $d_{x^2-y^2}$ for 
$J/t$ greater than $\sim 0.2$.

We would like to go back and discuss the fact that 
the spin fluctuation operator can break the string connecting
the two holes. These processes, however, compete with 
processes which lower the pair hole energy discussed previously 
by allowing the pair to delocalize when the connecting string is present 
and with other processes which accommodate the hole motion of each
hole within the string. Namely, when one hole creates 
a string, another hole near the first hole, has increased probability
to follow the string created by the first hole, thus, they are experiencing 
pairing correlations. Therefore, on variational grounds
we believe that the pairs of holes, depending on the parameter regime, 
may have string-mediated pairing correlations.

Here, we would like to discuss what happens when the level of
doping increases. As it has been discussed earlier in this paper, 
in order for the string-like correlations to be effective, we do not need to
require long range antiferromagnetic(AF) order. This is because
the cuprates are described by an intermediate parameter regime,
 where $t/J\sim 3$.
In this intermediate regime, we found that strings of length only three 
lattice spacings can
reproduce the single hole dispersion reasonably well (see Fig.~\ref{fig10}).
Therefore, short-range  {\it string correlations} only require that 
AF spin-spin correlations exist up to such rather short distance of two 
or three lattice spacings away  from the hole. 
This requirement is believed to be true even at optimal doping.
Hence, we believe that such short-range AF spin correlations,
required for the string mediated interaction between pairs of holes, 
could still be present in the cuprates even at optimum doping.

\section{conclusions}
Our findings suggest a new framework to understand doped 
quantum antiferromagnets and possibly the pairing mechanism
in the cuprate superconductors.

First, we illustrated that the high resolution ARPES studies\cite{ronning,graf}
from cuprates suggest the following.

(a) The hole in the doped quantum
antiferromagnet becomes a rather well-defined
quasiparticle and it is always attached to a {\it string} of
overturned spins which follows the hole in its
motion via  processes of coherent pair spin-flip  which lead to 
a reassignment of the hole birth site. 

(b) The high energy features and dispersion discussed in this
paper, which are seen in ARPES, can be understood as 
 ``internal'' excitations of the ``spin-polaron''. In this paper, it is shown
that these excitations
are responsible for the vanishing of the quasiparticle from the
lowest string state at the $\Gamma$ point and for the transfer of
spectral weight to higher energy string excitations. In addition,
they may be responsible for the high energy anomalous
dispersion found in the most recent ARPES studies\cite{ronning,graf}.

Second, a simple model is postulated where only states of holes connected with 
unbroken strings of overturned spins are retained.
This model gives a very good quantitative account of both the 
quasiparticle dispersion and the higher energy string excitations.

Third, motivated by both the fact that our simple model 
of string excitations can explain the results of the 
$t-J$ model and the fact that the new features of ARPES studies can be 
interpreted as arising from such string excitations, we further
consider the role of such string excitations in the pairing 
mechanism. We show that pairs of holes experience attractive
pairing interaction due to exchange of such string excitations.
Namely, when one of the holes creates such a string of overturned
spins in order to be able to move, another hole takes advantage of the
created path of overturned spins and follows it, thus, restoring
part of the destroyed antiferromagnetic order. 

The proposed pairing mechanism could be superior to those of models
which are based on exchange of antiferromagnetic spin fluctuations
for the following two main reasons:
(i) As we discussed in the previous section, 
our model does not require long range antiferromagnetic order
but rather short-range antiferromagnetic spin correlations.
(ii) The pairing energy scale is much larger than the typical
energy scale that a spin-fluctuation exchange model would
give, since the effective attraction in our string-exchange
model arises from exchange of short-range high energy excitations.


\section{Acknowledgments}
I would like to thank P. Coleman and V. Dobrosavljevic for 
discussions. This work was supported by NASA under grant No NAG-2867.

\end{document}